\documentclass[letterpaper,10pt]{article} 

\usepackage{opticameet3} 

\newcommand\authormark[1]{\textsuperscript{#1}}

\usepackage{amsmath,amssymb}
\usepackage[colorlinks=true,bookmarks=false,citecolor=blue,urlcolor=blue]{hyperref} 
\usepackage{acronym} 

\begin{document}

\acrodef{CU}{Centralised Unit}
\acrodef{DU}{Distributed Unit}
\acrodef{FSO}{Free Space Optics}
\acrodef{MEC}{Multi-Access Edge Computing}
\acrodef{MPLC}{Multi-Plane Light Conversion}
\acrodef{ORAN}{Open Radio Access Network}
\acrodef{PON}{Passive Optical Network}
\acrodef{TWDM}{Time and Wavelength Division Multiplexing}
\acrodef{UPF}{User Plane Function}

\title{Connected OFCity Challenge: an updated perspective on technology for connected cities}


\vspace{-5mm}

\author{Marco Ruffini,\authormark{1} Chongjin Xie,\authormark{2}, Lei shi,\authormark{3} and Jun Shan Wey \authormark{4}}

\address{\authormark{1} CONNECT centre, School of Computer Science and Statistics, Trinity College Dublin, Ireland \\
\authormark{2}Alibaba Infrastructure Service, Alibaba Cloud, NY, USA; \authormark{3}Damo Academy, Alibaba Cloud, Beijing, China\\
\authormark{4}Verizon Communications, Bellevue, Washington 98005, USA\\}

\email{\authormark{}marco.ruffini@tcd.ie, chongjin.xie@alibaba-inc.com, jun.shan.wey@verizon.com} 

\vspace{-4mm}
\begin{abstract}
This paper gives an update on technologies discussed during three OFC events, called 'The Connected OFCity Challenge', from 2016 to 2018. It focuses on research development and field deployment of Passive Optical Networks and Cloud-Based technologies.
\end{abstract}


\section{Introduction}
OFCity was a new type of event organised at the annual Optical Fiber Communications conference in 2016. It was designed as a workshop with the scope of discussing use of communications technology to develop a future smart/connected city, focusing especially on optical access network technology \cite{OFCity2016}. Following its success, the event was organised for two additional years. In 2017, the focus moved towards the development of technology for target use cases, such as support for a remote networked musical performance, support for AR/VR applications and for autonomous vehicles, with a strong focus on low latency and reliability \cite{OFCity2017}. The third and final event, in 2018, focused on technology to deliver broadband to rural areas in the developing world \cite{OFCity2018}, to provide connectivity at low cost, in harsh environment and with unreliable supplies (e.g, limited and intermittent energy supply).

Across the three editions of OFCity, we assisted to the proposal of several innovative solutions, spanning areas form networking to applications and addressing urban and rural areas, from developed and developing world. This article provides an update on some of the technologies proposed during the three OFCity events, at a distance of more than 5 years, to discuss how some of the ideas have progressed in the research community, in standardisation and field deployment. 

The following areas are the focus of our discussion: the evolution on architectures, deployment and usage of Passive Optical Networks (PONs); and the development of broadband for the developing world based on data-centre oriented architectures.



\section{PON evolution: from residential services towards pervasive usage}

\subsection{PON field deployment beyhind residential services}
PONs are considered one of the most prominent technologies to address scalability at high capacity, i.e., cost vs. number of subscribers. A recurring theme across the three OFCity events was the use of multi-wavelength PONs as a solution to provide ubiquitous multi-tenant connectivity to any type of service. These services range from residential access up to back-/fronthauling of fiber-enabled dense mobile cells. Support for fixed wireless convergence, mobile edge computing, sofwarisation and flexibility was also envisioned.

Fast forward to present day, many of the visions deliberated by the OFCity teams have been realized. For example, the multi-wavelength NG-PON2 (four wavelengths each at 10Gb/s line rate) has been deployed by Verizon and enables the convergence of mobile connectivity, residential, and business services into one PON infrastructure. For example, in Verizon’s Intelligent Edge Network (iEN) architecture, NG-PON2 optical line terminal (OLT), 4G baseband unit (BBU), and 5G centralized radio access network (CRAN) equipment all share an access site to provide mobile wireless, fixed wireless access (FWA), fiber-to-the-x (home, building, cell site), and point-to-point fiber connectivity. 
For single-wavelength PON systems, the 10Gb/s symmetric PON (XGS-PON) has begun mass deployment, mainly for residential, business, and backhaul. According to Dell’Oro, XGS-PON accounted for 15\% of the overall PON market in 2021. It is expected to reach 55\% by 2026 and could overtake GPON in 2024. Paving the way for future upgrade, higher-speed PON systems for beyond 10Gb/s are being standardized in the ITU-T, inter alia, single wavelength 50G-PON, multi-wavelength 50G-TWDM-PON, and WDM-PON at 25 Gb/s per wavelength.
    


Another topic debated in all three-year’s OFCity events is the adoption of coherent technology in access networks. On this, the jury is still out. A coherent UDWDM-PON was proposed during the NG-PON2 standardization stage \cite{UDWDM}. While the standard has reserved part of the L-band for point-to-point WDM overlay, it does not specifically call for the use of coherent technology. Recently, propelled by advancements in digital signal process, there is renewed interest in coherent technology for access. Standardization for a point-to-multipoint coherent optical subcarrier aggregation technology capable of a minimum of 25 Gb/s per wavelength for metro-access network applications is being discussed in the ITU-T \cite{G.Sup.VHSP,OpenXR}.

Fixed-Wireless Access (FWA) technology was briefly discussed in the OFCity events. FWA provides broadband services to customers by connecting customer premise equipment to wireless network, essentially bringing 5G coverage to inside customer premises. It has turned out to be the fastest growing broadband service offering especially in urban and dense urban areas. For example, the total number of Verizon’s FWA subscribers is expected to grow from  about 100,000 in 2021 to up to 5 million in 2025. FWA can be supported by either PtP or PON infrastructure.\cite{WeyECOC}

\subsection{PON for low latency and MEC connectivity}
Low latency is becoming an increasingly important requirement for using PON technology for 5G and future 6G networks. It is required both to support \ac{ORAN} functional split and for low latency applications. These use cases have different requirements, especially on the placement of the different virtual network functions or applications. The use of \ac{MEC} is considered a key architectural innovation for reducing latency, whether it is for hosting the \ac{DU} of an {\ac{ORAN}} system or also \ac{CU} and \ac{UPF} to run a low-latency application locally. \acp{PON} technology can in principle be used to provide connectivity to \ac{ORAN} and \ac{MEC} nodes, especially when used with multi-wavelength technology (either in \ac{TWDM} mode, or more simply as wavelength overlay) \cite{Mondal}. However, the tree topology of a PON can only support NORTH-SOUTH communications, which forces \ac{MEC} nodes to be located at the central office, increasing the overall latency. The idea of inter-ONU communication, enabling EAST-WEST traffic in a \ac{PON}, has been around for some time \cite{SpringerBookChapter29}, and was recently reconsidered as an option to interconnect MEC nodes \cite{PONMeshJOCN}. The architecture is characterised by a trade-off between full flexibility but high loss, for solutions that go through splitters multiple times \cite{PONMeshJOCN}, and lower loss but diminished flexibility, for solution that use splitters and diplexer in a fixed topology \cite{SpringerBookChapter29}. While the tradeoff remains, recent work \cite{PONMeshNet} showed that optical amplification can be avoided in the first-stage splitter nodes in most cases. In addition, new splitter technology based on \ac{MPLC} \cite{SpringerBookChapter29} could provide the best tradeoff between optical loss and flexibility.  

\subsection{PON disaggregtion}
Another interesting evolution in broadband network has been the progressive virtualisation of network functions, which involves almost every aspect, from packet switching, to radio and optical access. The benefits are manifold, from providing fine control in multi-tenant environment, to facilitating network slicing and development of third party intelligent control algorithms. With respect to PONs, virtualisation has evolved enormously since its first appearance around 2015 with the Central Office Reacrhitected as a Data Centre (CORD) ONF project \cite{CORD}. Innovations have spanned from softwarisation of the DBA algorithm for upstream transmission \cite{FASA}, to architectures enabling fine control of capacity allocation in multi-tenant environments \cite{vDBAJOCN}. These concepts have been brought already to standardisation \cite{BBF370i2}, \cite{BBF402} with some functionalities under development towards commercial solutions.

\subsection{Other PON applications}
Several other applications based on PONs were considered. For example, the Fiber-to-the-Room (FTTR) technology is a concept introduced for rural use cases, but has gained attention in recent years for home broadband. It is currently being standardized in the ITU-T for increasing broadband coverage at customer premise environment. It is essentially a very short reach indoor PON system, e.g., for factories and homes. 
    
Another interesting development is the integration of \ac{FSO} to support both satellite and terrestrial communications. New technological improvement in use of automatic tracking \cite{AutoTracking} to reduce pointing errors and use of adaptive optics \cite{FSOAdaptive} to mitigate the phase distortions induced by turbulence, are improving FSO reliability considerably, making it a good candidate to support a wide range of applications, including ORAN connectivity, MEC integration, etc. Dynamic steering can also be used to improve reliability under adverse weather condition. For example creating mesh networks that can re-route FSO signals through multiple hops over shorter distance (thus improving received signal power). 
New architectures are thus emerging, where hybrid PON-FSO systems are developed  \cite{hybrydFSO} with the aim of reducing fiber deployment cost for some difficult to reach access point locations. 
    
\section{Cloud-Based Broadband technologies}
The last event for OFCity focused on technologies for the developing world. A solution that was originally proposed to provide cost-effective, high capacity to rural areas in the developing world, based on data-center and cloud technologies \cite{OFCity2018}, has now progressed to address the broader issue of cost-effective 5G deployment. 
In this architecture, the broadband services can be provided through edge cloud nodes, which are then connected to service hubs located in office buildings, schools, stores, or community centers to provide wireless or wireline access to end users. The edge nodes are connected to central cloud data centers so that the end users can access cloud services such as computation, storage and communications.

Data Center Interconnect (DCI) technologies developed in the past few years focus on hyperscale applications (for central cloud computing) and typically are designed for systems with tens of Tb/s capacity \cite{JOCN_Xie}, which are excessive for edge applications where only a few hundred Gb/s capacity is needed.  To solve the problem, an open whitebox optical transport system was developed specifically for edge applications \cite{ECOC_Zheng}. The system integrates transponders and optical line components in one RU (rack unit) compact box, and SONiC (Software for Open Networking in the Cloud), an open-source network operating system (NOS) based on Linux, is adopted as the equipment operating system. It consists of a few pluggable cards, including optical line component cards and transponder cards, where a transponder card with pluggable digital coherent (DCO) modules can provide a capacity between 100 and 400 Gb/s. The architecture is based on SONiC architecture, it reuses the Redis, PMON, and platform drivers, but it also adds optical abilities to the platform. It supports OpenConfig optical YANG models in RESTCONF and Telemetry container, and introduces Optical Linecard State Service (OLSS) and Linecard Abstraction Interface (LAI) to control different vendors’ linecards \cite{ECOC_Zheng}. OLSS can manage all kinds of optical components in the optical transport linecards.

Another important application of cloud-based broadband is the cost-effective deployment of 5G mobile communication networks, as low latency and wide connectivity are becoming a new socio-economic and living infrastructure. The use of 5G networks combined with the edge nodes and service hubs can provide more convenient and powerful network access to the surrounding areas, as well as create conditions for the further enhancement of their economies and lives. This extends the concept of cloud computing to 5G networks, turning them into 5G cloud services, lowering operation and maintenance thresholds for realizing inclusive 5G.
In the concrete implementation, 5G control plane elements have been deployed in the cloud, so that users can connect their edge devices with built-in base stations and core network forwarding elements through fiber, to access the control plane elements in the cloud. This combines a 5G network with large throughput and low latency for local use. With cloud computing technology, the turn-up time for the entire 5G network service is reduced to hours, and network operation and troubleshooting can be automated in the cloud. The cost of a 5G network supporting more than 1,000 users and 10G of traffic is less than \$50,000. Anyone and any enterprise organization can build a cost-effective high-speed network to get an inclusive 5G cloud service to meet the demands of local economic development.

\section*{Acknowledgments}
\footnotesize
We would like to acknowledge the organisers and participant to the three OFCity events that run during the OFC conference between 2016 and 2018. Results discussed in this work were also funded by EI grant DT 2019 0014B (Freespace), and SFI-17/CDA/4760, 18/RI/5721.

\end{document}